\documentclass{jfm}
\usepackage{color}
\usepackage{amstext}
\usepackage{amssymb}
\usepackage{graphicx}
\usepackage{amsmath}
\usepackage{bm}
\usepackage[mathscr]{euscript}
\usepackage{amsfonts}

\usepackage{tcolorbox}
\newtcolorbox{mybox}[1]{colback=red!5!white,colframe=red!75!black,fonttitle=\bfseries,title=#1}

\newcommand\const{\mathrm{const}}

\newcommand\vA{\boldsymbol{A}}

\newcommand\vX{\boldsymbol{X}}
\newcommand\vP{\boldsymbol{P}}

\newcommand\vV{\boldsymbol{V}}

\newcommand\vx{\boldsymbol{x}}

\makeatletter
\def\@makefnmark{\hbox{\@textsuperscript{\normalfont\@thefnmark}}}
\makeatother

\begin{document}

{\title[A theory of flying/swimming saucers.]
{A theory of flying/swimming saucers. Exact solutions for rectilinear locomotion.}}

\author[V. A. Vladimirov]
{V.\ns A.\ns V\ls l\ls a\ls d\ls i\ls m\ls i\ls r\ls o\ls v$^{1,2,3}$}

\affiliation{$^1$ Sultan Qaboos University, Oman,\ $^2$ University of York, UK,\ $^3$ University of Leeds, UK}

\setcounter{page}{1}\maketitle \thispagestyle{empty}

\begin{abstract}

We study  self-propulsion (or locomotion) of a  robot (or an underwater vehicle) in an inviscid incompressible  fluid.
The robot's body is rigid, while its locomotion ability is due to an internal actuator,
which can perform controlled translational and rotational oscillations.
Our attention is focused on two classes of the plane analytic exact solutions, describing rectilinear locomotion.
Solutions of the first class describe the \emph{tumbling locomotion}, while
the second class corresponds to the \emph{zigzag locomotion} without tumbling.
We show, that  tumbling locomotion is more efficient.
Both classes of solutions show, that the use of actuator
allows to choose any desired direction and any speed of locomotion.
As a special case, we consider the self-propulsion caused by small-amplitude and high-frequency actuator oscillations.
The exact and elementary character of our solutions makes the results  potentially
useful as the tests to verify physical and engineering models, as well as
numerical and asymptotic results.
In contrary to many recent publications in this area,
the material is accessible to UG students in Engineering, Physics, and Applied Mathematics.

\end{abstract}

\section{Introduction}

This paper presents exact solutions of rectilinear locomotion (or self-propulsion) of an oscillating rigid  body in a fluid.
Studies of self-propulsion represent a high-impact and extremely active area of fluid dynamics,
where the majority of related papers are devoted to the locomotion in a viscous fluid (biological applications, micro- and nano-fluidics)
or in an inviscid fluid (biological and engineering applications).
For a viscous fluid see \emph{e.g.} \cite{Taylor, Lighthill, Childress, Childress1, Lauga, Pedley, Pedley1, Vladimirov1, Vladimirov2}
and reviews by \cite{Elgeti, Quillen}.
For an inviscid fluid see \cite{Benjamin, Saffman, Galper, Childress, Colgate, Sharma}.
The connection between these two directions is a subject of various physical and engineering models, as well as intense computations,
see \emph{e.g.} \cite{Childress, McHenry, Colgate, Childress2}.
Our paper is devoted to the most intriguing  case: the locomotion in an inviscid potential flow without any shedding of vorticity.
The first related example belongs to \cite{Benjamin, Benjamin1}.
\cite{Saffman} described a general mechanism of self-propulsion due to oscillations of  virtual (or added) mass.
An extension of this idea to a viscous fluid is given in \cite{Childress1}.
Saffman considered an ellipsoid with an eccentricity changing periodically with time.
Simultaneously, there was an internal mass (an actuator), oscillating along its main axis.
The conservation of momentum showed, that the averaged translational velocity of an ellipsoid was not zero.
Then, various aspects of the problem where developed by \cite{Lugovtsov, Galper, Kelly, Kozlov, Melli, Chambrion, Kilin1, Kilin2}.
The majority of related recent papers have a mathematical character, which require a rather advanced training for their understanding.
In contrary, the focus of our study is to analyse the rectilinear self-propulsion of an oscillating rigid body with the use of elementary exact analytic solutions.
We introduce a simple, but well-controlled robot.
Physically, the self-propulsion in our case is generated by oscillations of virtual mass, related to periodic oscillations of body's orientation,
which represents an extension of the original Saffman's idea.
Our solutions show that the robot can have any desired direction of rectilinear locomotion and a speed,
which is interesting for engineering applications.
Say, the robot of size $1 m$ can move with an averaged speed $\sim10 cm/s$ or faster.
The exact solutions can also be useful to validate the results, obtained by:  computations, physical and engineering modeling,
and the use of asymptotic methods.

In Section 2 we describe the construction of the robot and derive its equations of motion.
The robot consists of two pieces only: a rigid body and an actuator.
The latter represents a rigid body,
performing controlled translational and rotational oscillations.
Our derivation of equations is similar to that of \cite{Lamb, Benjamin, Saffman, Kozlov, Moffatt}.
We concentrate our study on the most instructive (and confusing for some researchers) case of the self-propulsion
with zero values of total momentum and total angular momentum.

In Section 3 we present two classes of exact analytic solutions.
The first class is the \emph{tumbling locomotion}, while
the second class is the \emph{zigzag locomotion}.
The latter name reflects the geometry of trajectory along with periodic oscillations in robot's orientation.
Technically, our results are based on the splitting of robot's motion into two parts:
an active motion with velocity $\vV_*(t)$, and inertial  oscillations $\vX_0(t)$.
The  main results are given as simple exact formulae for robot' velocity, and most importantly -- for its averaged velocity.
We show that the direction of this velocity and the speed of self-propulsion can be chosen arbitrarily.
Remarkably, the tumbling locomotion has no oscillations in $\vV_*(t)$, hence it appears as much more efficient
in terms of a ratio between the energies of oscillatory and averaged motions.
In the  zigzag locomotion the oscillatory
energy usually dominates over the locomotion energy.

Section 4 is devoted to the examples, illustrating  particular regimes of locomotion.
In Example 1 we present four particular cases of  exact solutions for simplified regimes of actuator's motion.
The simplest exact solution, presented in Example 2, may be seen as the most interesting for engineering applications.
Example 3 is devoted to clarifying physical and mathematical roles of initial conditions.
Here we also present the simplest exact solution from a different class.
In Examples 4\&5 we derive some results, related to small-amplitude robot's motions.
Here, we emphasise different possibilities of introducing small parameters into the problem,
including the case of Vibrodynamics, where small amplitude of oscillation is complimented by its high frequency \cite{Vladimirov0}.
This type of oscillations can be invisible for a naked eye,
hence such moving robot can look like a flying  saucer, when it flies or swims in a fluid without any visible reasons.
In particular we show, that the averaged speed of self-propulsion is proportional to $\varepsilon^2$, where $\varepsilon$ represents
an amplitude of actuator's vibrations. The self-propulsion velocity of order  $\varepsilon^2$ is typical for a viscous fluid,
see \emph{e.g.} \cite{Lighthill, Lighthill1, Childress, Pedley, Pedley1, Vladimirov1, Vladimirov2}, therefore
 the considered mechanism of self-propulsion is as efficient as that in a viscous fluid.
In Example 6 we give numerical estimations of averaged locomotion velocity,
which show that considered mechanism of locomotion may represent some engineering interest.
At the end of the paper, we make few remarks, devoted to possible generalisations, applications, and links to other areas.
\vskip -3mm
\begin{figure}
\centering\includegraphics[scale=0.8]{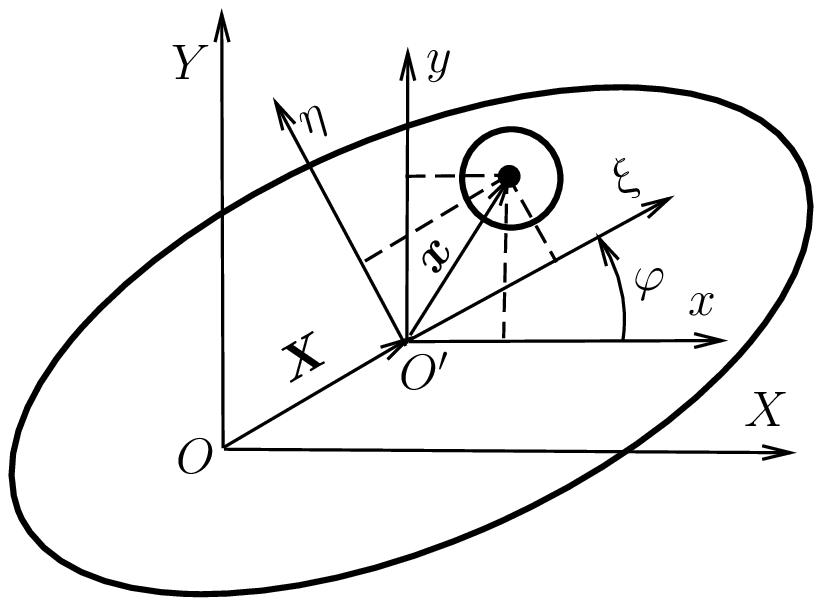}
\vskip -3mm
\end{figure}

\section{Robot's description and equations of motion}

The plane mechanical system (we call it `a robot') consists of a rigid body containing an actuator,
and moves in an infinite fluid.
A body is symmetric with respect to two mutually orthogonal axes (\emph{e.g.} it is an ellipse), its centroid is a point $O'$.
An actuator represents, say, a rigid disc, which relative motion can be precisely controlled.
A fluid is homogeneous, inviscid, and incompressible; its motion is irrotational with a state of rest at infinity.
Any external forces are  absent.
We use three systems of cartesian coordinates: $OXY$ is an inertial laboratory system, $O'xy$ is a moving system
with the axes  parallel to $OXY$, and $O'\xi\eta$ is a system, fixed with a body; with directions
$\xi$ and $\eta$ coincide with the axes of body's symmetry, see the figure.
The mass and  moment of inertia of a body are  $M^b$ and $J^b$.
We denote $M_1\equiv M^b+\mu_1$, $M_2\equiv M^b+\mu_1$, and $J^{bf}\equiv J^b+J^f$, where $\mu_1$, $\mu_2$ are the virtual masses of fluid
in the $\xi,\eta$-directions and $J^f$ is a virtual moment of fluid inertia.
An actuator mass is $m$, its moment of inertia is $j$.
Motion of a body is described by a vector  $\vX(t)\equiv\overrightarrow{OO'}$
and by an angle $\varphi(t)$ between the $x$ and $\xi$ axes, where $t$ is time.
The location of the actuator's centre of mass is  $\vX(t)+\vx(t)$.
We use following notations
{\setlength{\arraycolsep}{2pt}
   \renewcommand{\arraystretch}{0.8}
\begin{eqnarray}\label{matrix1}
&& \vX(t)=\begin{bmatrix}X\\Y\\ \end{bmatrix},\
 \vV^b(t)\equiv\dot\vX=
  \begin{bmatrix}U\\V\\ \end{bmatrix}_{xy}=
 \begin{bmatrix}u\\v\\ \end{bmatrix}_{\xi\eta}
 =\left\{\mathbb{\vA} \begin{bmatrix}u\\v\\ \end{bmatrix}\right\}_{xy}\\
 &&\vx(t)=
 \begin{bmatrix}x\\y\\ \end{bmatrix}=\left\{\mathbb{\vA}\begin{bmatrix}\xi\\ \eta\\ \end{bmatrix}\right\}_{xy}=
\begin{bmatrix} \xi\\ \eta\\ \end{bmatrix}_{\xi\eta},\quad \mathbb{\vA}(\varphi)\equiv
\begin{bmatrix}\cos\varphi& -\sin\varphi\\ \sin\varphi&\cos\varphi\\\end{bmatrix}\label{matrix2}\\
&&\vV^d(t)=\dot\vX+\dot\vx, \quad
\dot\vx=\begin{bmatrix}\dot x\\ \dot y\\ \end{bmatrix}_{xy}=
\left\{\mathbb{A}\begin{bmatrix} \dot\xi\\ \dot\eta\\ \end{bmatrix}
+\dot\varphi\mathbb{A}'\begin{bmatrix} \xi\\ \eta\\ \end{bmatrix}\right\}_{xy}\label{matrix3}
 \end{eqnarray}}
 where dots above letters stand for time derivatives,  $\mathbb{\vA}$ is a rotation matrix,
 $\mathbb{A}'\equiv d\mathbb{A}/d\varphi$; subscripts $xy$ and $\xi\eta$
denote $xy$ and $\xi\eta$-components of vectors correspondingly; the meaning of components without subscripts
are clear from a context; $\vV^d(t)$ is a velocity of actuator's center of mass,
its $\xi\eta$-projections follow after multiplication \eqref{matrix3} by $\mathbb{A}^{-1}$ and transformations:
\begin{equation}\label{matrix4}
\vV^d(t)={\setlength{\arraycolsep}{2pt}
   \renewcommand{\arraystretch}{0.9}
\begin{bmatrix} u+\dot\xi-\eta\dot\varphi\\ v+\dot\eta+\xi\dot\varphi\\ \end{bmatrix}_{\xi\eta}}\
\end{equation}

The total momentum $\vP^{\text{total}}$  is conserved with time due to the absence of external forces
\begin{equation}\label{momentum}
{\setlength{\arraycolsep}{2pt}
   \renewcommand{\arraystretch}{0.8}
\vP^{\text{total}}=\vP^{bf}+\vP^d=\const,\ \vP^{bf}=   \begin{bmatrix}M_1 u\\M_2 v\\ \end{bmatrix}_{\xi\eta},\
\vP^d=m\vV^d}
\end{equation}
where  the expression for momentum of `body+fluid' $\vP^{bf}$ is a classical result, see \cite{Lamb,Batchelor}.  Using \eqref{matrix4} we write
  \begin{equation}\label{momentum1}
  {\setlength{\arraycolsep}{2pt}
   \renewcommand{\arraystretch}{0.8}
 \vP^{\text{total}}=   \begin{bmatrix}(m+M_1) u+m(\dot\xi-\eta\dot\varphi)\\(m+M_2) v+m(\dot\eta+\xi\dot\varphi) \\ \end{bmatrix}_{\xi\eta}}
\end{equation}
Let $\vP^{\text{total}}\equiv 0$. Then \eqref{momentum1} yields
\begin{equation}\label{uv0}
{\setlength{\arraycolsep}{2pt}
   \renewcommand{\arraystretch}{0.8}
\vV^b (t)=\begin{bmatrix} u\\ v\\ \end{bmatrix}_{\xi\eta}=
\begin{bmatrix}\lambda_1 (\eta\dot\varphi-\dot\xi)\\ -\lambda_2(\xi\dot\varphi+\dot\eta) \\ \end{bmatrix},\ \text{where}\
 \lambda_i\equiv m/(m+M_i),\ i=1,2}
\end{equation}
Next, we express  components of $\vV^b$   in $xy$-axes:
\begin{equation}\label{uv}
{\setlength{\arraycolsep}{2pt}
   \renewcommand{\arraystretch}{0.8}
\vV^b=\begin{bmatrix} U\\ V\\ \end{bmatrix}_{xy}=\left\{\mathbb{A}\begin{bmatrix} u\\ v\\ \end{bmatrix}\right\}_{xy}=
\left\{\mathbb{A}\begin{bmatrix}\lambda_1 (\eta\dot\varphi-\dot\xi)\\ -\lambda_2(\xi\dot\varphi+\dot\eta) \\ \end{bmatrix}\right\}_{xy}}
\end{equation}
Then, we transform \eqref{uv}  with the use of explicit form of $\mathbb{A}$, and split it into two parts
\begin{eqnarray}
&&\vV^b (t)    = \dot \vX_{0}(t) +\vV_*(t) \quad
\text{or}\quad \frac{d}{dt} (\vX- \vX_{0})=\vV_*\label{XYdot}\\
&&{\setlength{\arraycolsep}{2pt}
   \renewcommand{\arraystretch}{0.8}
\vX_0=\begin{bmatrix}  X_{0}\\ Y_{0}\\\end{bmatrix}_{xy}\equiv -
\begin{bmatrix} \lambda_2\xi\cos\varphi -\lambda_1\eta\sin\varphi\\ \lambda_2\xi\sin\varphi+\lambda_1\eta\cos\varphi\\ \end{bmatrix}_{xy}=
-\Lambda\xi\begin{bmatrix} \cos\varphi\\  \sin\varphi\\ \end{bmatrix}_{xy}-\lambda_1\vx \label{XYdot1}}\\
&&{\setlength{\arraycolsep}{2pt}\renewcommand{\arraystretch}{0.8}
\vV_*=\begin{bmatrix} U_*\\  V_*\\ \end{bmatrix}_{xy}\equiv
\Lambda\begin{bmatrix} \dot\xi\cos\varphi +\dot\eta\sin\varphi\\ \dot\xi\sin\varphi-\dot\eta\cos\varphi\\ \end{bmatrix}}_{xy},
\quad \Lambda\equiv \lambda_2-\lambda_1\label{XYdot2}\\
&&\vX^2_0=(\lambda_2\xi)^2+(\lambda_1\eta)^2<\xi^2+\eta^2, \quad \vV^2_*=\Lambda^2(\dot\xi^2+\dot\eta^2)\label{inertial}
\end{eqnarray}
We call $\vX_0(t)$ an inertial displacement, and $\vV_*(t)$ - a velocity of active motion.
The inequality in \eqref{inertial} is valid since $\lambda_1<1,\, \lambda_2<1$ by \eqref{uv0}.
It shows that the displacement $\vX=\vX_0(t)$ is localized inside the circle, smaller than a robot's size,
hence this part of motion does not contain any locomotion.
Physically, it is similar to a low of constant velocity of a centre of mass.
Indeed, for a circular body $\Lambda=0$ and $\vX+\lambda_1\vx=0$, which confirms this similarity.
The second equality in \eqref{inertial} demonstrates a surprising link between $\vV_*$
and the actuator's translational oscillations.
It immediately imposes an upper bound on the speed of possible locomotion.

The results \eqref{XYdot}-\eqref{XYdot2} still contain an unknown function $\varphi(t)$, which can be obtained from
the conservation of a $z$-component of total angular momentum
\begin{equation}\label{L1}
 L^{\text{total}}=(\vX\times \vP^{bf})_z+ ((\vX+\vx)\times \vP^d )_z+ J^{bf} \dot\varphi+{j}(\sigma+\dot\varphi)=\const
\end{equation}
where $\sigma(t)$ is an angular velocity of actuator in the $O'\xi\eta$-system. We choose $L^{\text{total}}=0$.
Then the substitution of $\vP^d=\vP^{\text{total}}-\vP^{bf}$ (where also $\vP^{\text{total}}=0$) yields
\begin{equation}\label{L2}
 L^{\text{total}}=-(\vx\times \vP^{bf})_z+J\dot\varphi+{j} \sigma=0;\quad J\equiv J^{bf}+{j}
\end{equation}
The use of \eqref{matrix2},\eqref{momentum}, \eqref{L2} and straightforward transformation involving the explicit form  of $\mathbb{A}$, give
\begin{eqnarray}
&&-(\vx\times \vP^{bf})=-xP_2+yP_1= -M_2 v\xi+M_1 u\eta\nonumber\\
&& L^{\text{total}}=-M_2 v\xi+M_1 u\eta+J\dot\varphi+{j} \sigma=0\label{L3a}
\end{eqnarray}
Now,  the substitution of $u,v$ from \eqref{uv0} into \eqref{L3a} produces the required equation for $\varphi(t)$
\begin{eqnarray}\label{dotphi}
\dot\varphi=\Omega(t)\equiv\frac{\delta_1\dot\xi \eta- \delta_2\xi \dot\eta-{j} \sigma}{J+ \delta_1\eta^2+\delta_2\xi^2},\quad \delta_1\equiv
\lambda_1 M_1,\ \delta_2\equiv\lambda_2 M_2
\end{eqnarray}
Equations \eqref{XYdot}-\eqref{XYdot2}, \eqref{dotphi} give a full description of robot's motion, provided that  functions $\xi(t),\eta(t),\sigma(t)$
are prescribed.
We do not restrict ourselves with any particular initial condition,
the related clarification will be given it the Example 3 of Section 4.
Apparently, there are infinitely many related solutions.
Our aim is to find some simple exact solutions for the rectilinear locomotion.
We simplify the problem by considering only time-periodic control functions $\xi(\tau),\eta(\tau),\sigma(\tau)$,
which, (according to a well-known theorem) produce only time-periodic solutions.
Each such function has the following properties:
(i) it is $2\pi$-periodic in $\tau\equiv\omega t$, where $\omega=\const$ is a frequency;\
(ii) it has an average, defined as
$\overline{f}\equiv \langle {f}\,\rangle \equiv \frac{1}{2\pi}\int_{\tau_0}^{\tau_0+2\pi}
f(\tau)\, d \tau $ for any $\tau_0=\const$,
where a double notation $\overline{f}\equiv \langle {f}\rangle$ is used for avoiding cumbersome formulae;\
(iii) it can be split into the averaged and oscillating parts $f(\tau)=\overline{f}+\widetilde{f}(\tau)$,
where $\langle \widetilde f\, \rangle =0$
and $\overline{f}\equiv \langle {f}\rangle=\const$.
Then  $\vV_*$ is expressed in the $xy$-axes as:
{\setlength{\arraycolsep}{2pt}\renewcommand{\arraystretch}{0.8}
\begin{eqnarray}
&&\vV_*=\begin{bmatrix} U_*\\  V_*\\ \end{bmatrix}=
\omega\Lambda\begin{bmatrix}&\cos\varphi\ &\sin\varphi\\ &\sin\varphi\ -&\cos\varphi\\ \end{bmatrix}
\begin{bmatrix} \widetilde{\xi}_{\tau}\\  \widetilde{\eta}_{\tau}\\ \end{bmatrix}=
\omega\Lambda\begin{bmatrix} 1 & 0\\  0 &-1\\ \end{bmatrix}
\mathbb{A}^{-1}(\varphi)
\begin{bmatrix} \widetilde{\xi}_{\tau}\\  \widetilde{\eta}_{\tau}\\ \end{bmatrix}\label{V-tilde}\\
&&\text{where}\quad \varphi_{\tau}=
\frac{\delta_1\widetilde{\xi}_{\tau} \widetilde{\eta}_{\tau}- \delta_2\widetilde{\xi}_{\tau} \widetilde{\eta}_{\tau}-j
\sigma/\omega}{\delta_1\widetilde{\eta}^2+\delta_2\widetilde{\xi}^2+J}\label{phi-tilde}\\
&&\text{and}\quad \Lambda\equiv \frac{m(\mu_1-\mu_2)}{(m+M_1)(m+M_2)},\ \delta_i\equiv\frac{m M_i}{m+M_i},\ M_i\equiv M+\mu_i,\quad i=1,2\label{Lambda}
\end{eqnarray}}
where three arbitrary control functions are:
\begin{equation*}
  \xi=\widetilde{\xi}(\tau), \quad \eta=\widetilde{\eta}(\tau), \quad \sigma=\sigma (\tau);\quad \tau\equiv\omega t
\end{equation*}
In the above text all the functions and variables are dimensional.
However, they can be also seen as dimensionless, which can be achieved by the use of following characteristic parameters:
mass $M_{\text{char}}\equiv M_0+m$, length $L_{\text{char}}\equiv \max(|\xi|, |\eta|)$,
and  time $T \equiv 1/\omega$.

\section{Exact solutions for the rectilinear locomotion}

We obtain exact solutions by an \emph{inverse method}:
we propose simple forms of $\widetilde{\xi}(\tau), \widetilde{\eta}(\tau)$ and $\varphi(\tau)$,
after that the required motion of an actuator $\sigma (\tau)$ is calculated from  \eqref{phi-tilde}.

For the \textbf{\emph{tumbling  locomotion}} we introduce the control functions as
\begin{eqnarray}\label{contr-funct-1}
{\setlength{\arraycolsep}{2pt}
   \renewcommand{\arraystretch}{0.8}
\begin{bmatrix}\widetilde{\xi}\\ \widetilde{\eta}\\ \end{bmatrix}=\mathbb{A}(\varphi)\begin{bmatrix}\xi_0\\ \eta_0\\ \end{bmatrix};\quad
\varphi(\tau)=-\tau,\ \overline{\Omega}=-\omega,\ \widetilde{\Omega}\equiv 0;
\quad \vx=\widetilde{\vx}=\mathbb{A}^2(\varphi)\begin{bmatrix}\xi_0\\ \eta_0\\ \end{bmatrix}}
\end{eqnarray}
where $\xi_0$ and $\eta_0$ are constants.
There are four parameters possessing the dimension of frequency in \eqref{contr-funct-1}:
(i) $\omega$ is an oscillatory frequency of all control functions and solutions;
(ii) $\sigma$ is an angular velocity of an actuator, we call it  \emph{self-rotation};
(iii) $\sigma^{\text{orb}}$ is an angular velocity of the actuator's center about the centroid $O'$, we call it an \emph{orbital rotation};
and (iv) $\Omega$ is an angular velocity of the robot's body.
Both $\sigma$ and $\sigma^{\text{orb}}$ are defined in the rotating $O'\xi\eta$-coordinates.
Formulae \eqref{contr-funct-1} mean that the robot's angular velocity $\overline{\Omega}=-\omega$, and $\sigma^{\text{orb}}=-\omega$.
The last expression in \eqref{contr-funct-1} is obtained with the use of \eqref{matrix2},
it shows that the actuator's orbital angular velocity in $Oxy$-system is $-2\omega$.
The substitution of \eqref{contr-funct-1} into \eqref{V-tilde}-\eqref{Lambda} and transformations yield
\begin{eqnarray}\label{V-tumbl}
&&{\setlength{\arraycolsep}{2pt}\renewcommand{\arraystretch}{0.8}
\vV_*=\overline{\vV}_*=\omega\Lambda \begin{bmatrix} \eta_0\\ \xi_0\\ \end{bmatrix}_{xy}},\quad \widetilde{\vV}_*\equiv 0\\
&& \sigma(\tau)=\omega (J+2\nu(\tau))/j , \quad
\nu(\tau)\equiv \delta_1\widetilde{\eta}^2+\delta_2\widetilde{\xi}^2=\nu_0+\nu_1\cos 2\tau+\nu_2 \sin 2\tau\nonumber\\
&& \nu_0\equiv (\xi^2_0+\eta^2_0)\delta^+/2,\quad \nu_1\equiv (\xi^2_0-\eta^2_0)\delta^-/2, \quad \nu_2\equiv \xi_0\eta_0\delta^-,
\quad \delta^\pm\equiv\delta_2\pm\delta_1 \nonumber
\end{eqnarray}
One can see, that the robot robot can perform a rectilinear locomotion with a constant speed $\vV_*$ in any desired direction.
At the same time \eqref{V-tumbl}  shows a surprising absence of oscillations.
This rectilinear locomotion is disturbed only by inertial oscillations $\vX_0(\tau)$ \eqref{XYdot1}.
Simultaneously, an actuator performs both self-rotation and orbital rotation,
where  constant self-rotation $\overline{\sigma}\neq 0$ can be eliminated if we allow  $L^{\text{total}}= j\overline{\sigma}\neq 0$.
The simplicity of \eqref{V-tumbl} makes the tumbling locomotion  attractive for practical use in engineering devices.
However, the robot's tumbling may create various operational difficulties, say, in observations and in manipulations.
In order to avoid that, we study the \textbf{\emph{zigzag  locomotion}} where we eliminate any tumbling by imposing conditions
\begin{eqnarray}\nonumber
\overline{\varphi}=\overline{\Omega}=0,\quad \text{hence}\quad\ \varphi=\widetilde{\varphi},\
\Omega=\widetilde{\Omega}=\omega\widetilde{\varphi}_\tau,
 \end{eqnarray}
which also mean that the averaged orientations of the body's main axes are chosen coinciding with the $X,Y$-directions.
For simplifying the control, we allow an actuator oscillations only along a fixed internal rail of a fixed slope $\psi$
\begin{equation}\label{tilde}
  \widetilde{\xi}= \widetilde{\zeta}(\tau)\cos\psi,\ \widetilde{\eta}\equiv \widetilde{\zeta}(\tau)\sin\psi;
\end{equation}
which means that we reduce two arbitrary functions $\widetilde{\xi}(\tau)$ and $\widetilde{\eta}(\tau)$ to a single one $\widetilde{\zeta}(\tau)$.
Then, \eqref{XYdot2}, \eqref{dotphi} are reduced to
\begin{eqnarray}
&&{\setlength{\arraycolsep}{2pt}
   \renewcommand{\arraystretch}{0.8}
\vV_*= \overline{\vV}_*+\widetilde{\vV}_*=
\omega\Lambda\widetilde{\zeta}_\tau
\begin{bmatrix} \cos(\psi-\widetilde{\varphi})\\ \sin(\psi-\widetilde{\varphi})\\ \end{bmatrix}_{xy}
\label{UV}}\\
&&{j}  \sigma/\omega= \kappa_1 \widetilde{\zeta}\,\widetilde{\zeta}_\tau-\widetilde{\varphi}_\tau[J+
\kappa_2\widetilde{\zeta}^2],\quad 2\kappa_1\equiv \delta^-\sin 2\psi,\ 2\kappa_2\equiv\delta^+ -\delta^-\cos 2\psi
\label{dotphi1}
 \end{eqnarray}
The average rotation of an actuator vanishes  $\overline{\sigma}=0$ if $\langle\widetilde{\varphi}_\tau \widetilde{\zeta}^2\rangle=0$,
such a choice allows a further simplification of the control functions (Note: one can make $\overline{\sigma}\equiv0$
by allowing $L^{\text{total}}\neq 0$, without any changes in all other results).
We are looking for the exact solutions in the form
\begin{eqnarray}\label{solution-form}
\widetilde{\varphi}(\tau)=\widehat{\varphi}\sin\tau,\quad
\widetilde{\zeta}_\tau(\tau)=\sum_{n=1}^{\infty}(a_n\cos n\tau+b_n\sin n\tau);\quad n=1,2,\dots
\end{eqnarray}
where $\widehat{\varphi}, a_n, b_n$ are constants.
Next, we substitute  \eqref{solution-form} into  \eqref{UV}-\eqref{dotphi1}, which (in a general form) produces a rather cumbersome expression.
However, there is a drastic simplification of an averaged velocity, after the
use of the following Fourier series, see \cite{Watson}, p.22
\begin{eqnarray}\label{Bessel}
 {\setlength{\arraycolsep}{2pt}
   \renewcommand{\arraystretch}{0.8}
\begin{bmatrix}\cos(\widehat{\varphi}\sin\tau)-\widehat{J}_0\\ \sin(\widehat{\varphi}\sin\tau)\end{bmatrix}=
2\sum_{n=1}^\infty \begin{bmatrix} \widehat{J}_{2n}\cos 2n\tau\\ \widehat{J}_{2n-1}\sin(2n-1)\tau \end{bmatrix}}
\end{eqnarray}
where $\widehat{J}_n\equiv J_n(\widehat{\varphi})$ are Bessel functions of the first kind.
The use of  the Parseval's equality leads to the following expressions for a constant averaged locomotion velocity
and to the distribution of energy between the averaged and oscillatory motions
{\setlength{\arraycolsep}{2pt}
   \renewcommand{\arraystretch}{0.8}
\begin{eqnarray}
&&\overline{\vV}_*=\begin{bmatrix}  \overline{U}_*\\   \overline{V}_* \\\end{bmatrix}=\omega\Lambda
\begin{bmatrix}  W_1\cos\psi + W_2\sin\psi\\ -W_1\sin\psi+W_2\cos\psi\\ \end{bmatrix}_{xy};\quad
\begin{bmatrix}  W_1\\   W_2 \\ \end{bmatrix}\equiv
\sum_{n=1}^\infty \begin{bmatrix}  a_{2n}\widehat{J}_{2n}\\ b_{2n-1}\widehat{J}_{2n-1} \\ \end{bmatrix}\label{V-final}\\
&&\overline{\vV}_*^2=\omega^2\Lambda^2 (W_1^2+W_2^2);\quad
\widetilde{\vV}_*^2=\omega^2\Lambda^2\sum_{n=1}^\infty (a_n^2+b_n^2)/2-W_1^2-W_2^2\label{energy}
\end{eqnarray}}
\noindent
where the second equality \eqref{energy} is obtained with the use of \eqref{inertial}.
Expressions $W_1$ and $W_2$ can consist of finite or infinite number of terms,
depending on our choice of coefficients in \eqref{solution-form}.
From \eqref{V-final} one can conclude, that the direction locomotion can be chosen arbitrarily.
Since  the coefficients $a_n, b_n$ can be also chosen arbitrarily, than $|\vV_*|$ can be as high as we wish.
For decreasing the oscillatory motion of the robot we  accept in all the examples below that all odd coefficients $a_{2n-1}=0$
and all even coefficients $b_{2n}=0$.
Such a choice of coefficients keeps the averaged locomotion unchanged.

Notice, that in the zigzag regime any change of locomotion direction of or its speed does not require any changes of the robot's body orientation.

\section{Examples:}
The expressions \eqref{V-final}, \eqref{dotphi1}, \eqref{solution-form} are very general, here we present some clarifying special examples.

\textbf{\emph{1. Four particular classes of the zigzag locomotion:}}\
 The rail's orientation along the main axes $\xi$ or $\eta$ leads to
{\setlength{\arraycolsep}{2pt}
   \renewcommand{\arraystretch}{0.8}
\begin{eqnarray}\label{rail-0pi}
\psi=0: \ \text{then}\  \overline{\vV}_*=\omega\Lambda\begin{bmatrix} W_1\\ W_2 \end{bmatrix}; \quad
\psi=\pi/2: \ \text{then}\  \overline{\vV}_*=\omega\Lambda\begin{bmatrix} W_2\\ -W_1 \end{bmatrix}
\end{eqnarray}}
\noindent One can see that the direction of $\overline{\vV}_*$ in one case can be chosen arbitrarily,
while the second direction is perpendicular to the first one. Both $W_1$ and $W_2$ can take any values.
Alternatively, for arbitrary $\psi$, when we take $W_1=0$ (by taking all $a_n=0$) or $W_2=0$ (by taking all $b_n=0$), then  for all $n$ we obtain
\begin{eqnarray}\label{rail-psi}
&& {\setlength{\arraycolsep}{2pt}
   \renewcommand{\arraystretch}{0.8}
W_1=0: \ \overline{\vV}_*=-\omega\Lambda W_2\begin{bmatrix} \sin\psi\\ \cos\psi \end{bmatrix}, \quad
W_2=0: \ \overline{\vV}_*=-\omega\Lambda W_1\begin{bmatrix} \cos\psi\\ -\sin\psi \end{bmatrix}}
\end{eqnarray}
In these two cases  the  directions of $\overline{\vV}_*$ are again mutually orthogonal,
in addition they are directly linked to the rail's orientation.
One can see that cases $\psi=0$ and $\psi=\pi/2$ in \eqref{rail-psi} show that the oscillations of an actuator along one main axis lead to
the locomotion in a perpendicular direction of another main axis.

\textbf{\emph{2. A simplest case of the zigzag locomotion:}}\
Only one nonzero term in \eqref{solution-form} yields
\begin{eqnarray}
&& {\setlength{\arraycolsep}{2pt}
   \renewcommand{\arraystretch}{0.8}
\widetilde{\zeta}(\tau)=b_1\sin\tau,\quad \overline{\vV}_*=\omega\Lambda b_1\widehat{J}_1 \begin{bmatrix} \sin\psi\\ \cos\psi
\end{bmatrix}}\label{V-sin}\\
&&2j\sigma(\tau)/\omega= -\widehat{\varphi}(2J+\kappa_2 b_1^2)\cos\tau+
\kappa_1 b_1^2\sin 2\tau+\kappa_2 b_1^2\widehat{\varphi}\cos 3\tau\label{sigma-sin}
 \end{eqnarray}
One can see that the direction of $\overline{\vV}_*$ still can be chosen arbitrarily by the choosing of rail's orientation.
The required actuator's angular velocity $\sigma=\widetilde{\sigma}(\tau)$ \eqref{sigma-sin} represents pure oscillations
of the first three harmonics, when $\overline{\sigma}\equiv 0$.
The maximum speed $\vert\overline{V}_*\vert\simeq 0.58 \omega\Lambda b_1$ takes place at
$\widehat{\varphi}\simeq 1.84$, see \cite{Watson}, p.669.
The ratio between the oscillatory and locomotion energies may be presented as
$\langle\widetilde{\vV}_*^2\rangle/\langle\overline{\vV}_*^2\rangle= (1/2\widehat{J}_1^2)-1$;
at $\widehat{\varphi}\simeq 1.84$ this ratio is about $0.16$.
Hence, one can see that the finite amplitude zigzag locomotion can be rather efficient.

\textbf{\emph{3. The role of initial conditions:}}\
One may notice that the solutions \eqref{solution-form} or \eqref{V-sin}, taken in the class of continuous functions,
do not correspond to the initial state of rest:
\begin{equation}\label{rest}
   U=V=\dot\xi=\dot\eta= \dot\varphi=\sigma=0   \quad \text{at}\quad t=0
\end{equation}
At the same time, in the physical interpretation,  self-propulsion is
a motion, which starts from the state of rest.
However, our choice $\vP^{\text{total}}\equiv 0$ and
$L^{\text{total}}=0$ makes the initial conditions \eqref{rest} unnecessary if we allow an impulsive start.
Indeed, if our robot is moving arbitrarily and an actuator abruptly stops, then any motion of the robot and a fluid
is instantly and completely halted.
Indeed, any motion of a rigid body with the zero values of above integrals is impossible.
Then, using the invariance $t\to -t$ one can reverse this process and see that any motion can abruptly appear from the state of rest
after an actuator instantly starts moving with finite speeds.
Hence, a difference between a smooth start of robot's locomotion or its impulsive start is not essential for our case,
both these regimes can naturally appear from a state of rest at $t=0$.
Nevertheless, let us also briefly describe the solution with a smooth starting.
In this case one may replace \eqref{solution-form} with
\begin{eqnarray}\label{solution-form1}
\widetilde{\varphi}(\tau)=\widehat{\varphi}\cos\tau,\quad \widetilde{\zeta}_\tau=\widehat{\zeta}(\cos\tau-\cos k\tau)
\end{eqnarray}
where $\widehat{\zeta}$ is a constant, and $k\geq 3$ is an odd integer.
Calculations show that in this case
\begin{eqnarray}\label{WJ}
&& {\setlength{\arraycolsep}{2pt}
   \renewcommand{\arraystretch}{0.8}
\overline{\vV}_*=\omega\Lambda \widehat{\zeta}(\widehat{J}_1-\widehat{J}_k)\begin{bmatrix} \sin\psi\\ \cos\psi \end{bmatrix}}
 \end{eqnarray}
 which is qualitatively similar to \eqref{V-sin}.
 One can find that in this case all the conditions \eqref{rest} are valid.

\textbf{\emph{4. Linearization in the robot's angle:}} \
For $\widehat{\varphi}\ll 1$ a linearized version of \eqref{V-final} represents a special interest,
since it can be used as a test example in asymptotic theories.
In this case  \eqref{V-final} yields:
\begin{eqnarray}\label{linear-phi}
&& {\setlength{\arraycolsep}{2pt}
   \renewcommand{\arraystretch}{0.8}
\overline{\vV}_*=\frac{1}{2}\omega\Lambda b_1 \widehat{\varphi}\begin{bmatrix} \sin\psi\\ \cos\psi \end{bmatrix},\quad
\widetilde{\zeta}(\tau)=b_1\sin\tau}
\end{eqnarray}
where we use $J_n(\widehat{\varphi})= (\widehat{\varphi}/2)^n/n!+O(\widehat{\varphi}^2)$.
We emphasise that \eqref{linear-phi} contains  only one Fourier coefficient $b_1$, while all other coefficients
of  \eqref{V-final} appear only in nonlinear approximations.
The amplitude $b_1$ in \eqref{linear-phi}  is still arbitrary.
The direction of locomotion and  the oscillations of  an actuator are the same as in previous example.
For correct comparison with asymptotic theories, we notice that another small parameter can appear in \eqref{linear-phi} as $\Lambda\ll 1$,
if the robot's shape is only slightly different from a circular cylinder.
In this situation one can take $\widehat{\varphi}\sim\varepsilon$ and $\Lambda\sim\varepsilon$ hence $|\vV_*|\sim\varepsilon^2$.
One may recall that the swimming velocity of order  $\varepsilon^2$ (where $\varepsilon$ is a small amplitude of oscillations of
a swimming machine or a living creature) is typical for a viscous fluid,
see \cite{Lighthill, Childress, Pedley, Pedley1, Vladimirov1, Vladimirov2}. It shows that
the considered inviscid mechanism of self-propulsion is at least as efficient as that in a viscous fluid.

\textbf{\emph{5. Linearization in the robot's angle and in amplitude of displacement:}}\
A linearization in both $\widehat{\varphi}\ll 1$ and $b_1\ll 1$  keeps $\overline{\vV}_*$  the same as
 \eqref{linear-phi}, while $\sigma(\tau)$ is drastically simplified:
\begin{eqnarray}\label{Gener-linear}
&& {\setlength{\arraycolsep}{2pt}
   \renewcommand{\arraystretch}{0.8}
\overline{\vV}_*=\frac{1}{2}\omega\Lambda b_1 \widehat{\varphi}\begin{bmatrix} \sin\psi\\ \cos\psi \end{bmatrix},\quad
\widetilde{\zeta}(\tau)=b_1\sin\tau,\ \widetilde{\sigma}(\tau)= \widehat{\sigma}\sin\tau};
\quad \widehat{\varphi}=-j\widehat{\sigma}/J\omega
 \end{eqnarray}
Here we again obtain the quadratic in small amplitudes locomotion, provided that the value of $\Lambda$ is not small.
An internal parameter $b_1\ll 1$ is not related to the shape of the robot, therefore this example represents a physically different
(from the previous one) asymptotic solution.

\textbf{\emph{6. A numerical estimation of the locomotion speed}} follows from \eqref{V-tumbl} and \eqref{V-sin}
where $|\overline{\vV}_*|\simeq C\omega\Lambda \max(|\xi|,|\eta|)$,  $C\sim 0.5-1$.
The estimation of $\Lambda$ \eqref{Lambda} may be obtained if we
accept that the robot' body is an ellipse with semi-axes $\alpha>\beta$.
It gives virtual masses  $\mu_1=\rho\pi \beta^2$ and  $\mu_2=\rho\pi \alpha^2$,
where $\rho=\const$ is the density of a fluid.
Apparently, the robot operates in the conditions of neutral buoyancy $M+m=\rho\pi\alpha\beta$.
Then $\Lambda=g(1-h)/(1+g)(1+h)$ where $g\equiv m/M$ and $h\equiv \alpha/\beta$.
Taking $g=1/2, h=2$ we get $\Lambda=-1/9$.
The choice of experimental parameters   $\omega=1\, Hz\simeq 6\, s^{-1}$, $\Lambda\simeq 0.1$, and $\max(|\xi|,|\eta|)\simeq 1\, m$ yields
$|\overline{\vV}_*|\simeq 0.6C\, m/s$, which for $C\sim 1$ represents an impressive result.
Certainly, for small oscillations (see Examples 5 and 6) the speed will be much lower.
The increasing of $\omega$ gives an opportunity for faster self-propulsion.
It is interesting, that the averaged locomotion speed does not depend on the size of the robot, as well as on the direction of propagation.

\subsection{Concluding Remarks}
 \underline{Controllability}.  The question of \emph{controllability} of the robot is solved above, since the use of a rectilinear trajectory
(or a trajectory, consisting of several rectilinear parts) allows a robot's locomotion from any point of space to any other point.
An interesting question of robot's \emph{stability} or stabilization , see \emph{e.g.} \cite{Woolsay}, remains open.\

 \underline{Three- dimensional generalizations.}
The above theory can be straightforwardly extended to a \emph{three-dimensional disc-shape robot},
which performs only plane motions.\

 \underline{ Interaction with boundaries.} An oscillating/vibrating body is usually \emph{attracted/repulsed by an external boundary} or by other moving bodies,
see\emph{ e.g.} \cite{Bjerknes, Sennitsky, Vladimirov0}.
In a viscous fluid it can also perform an efficient \emph{exchange in angular momentum}, see \cite{MoffattNew}.
Hence, for a practical use of an oscillating robot, one should take into account  its interactions with various boundaries,
which could be used as an advantage for some applications.\

 \underline{Separation and Wake.}  The issue of a \emph{flow separation and wake} for the considered periodic motions remains open.
One may assume, that the full control of the robot's shape allows to optimize a real flow structure.\

 \underline{Technical potential.}
Possible \emph{technical developments} may include underwater robots, torpedoes, remotely operated or autonomous underwater vehicles, up to submarines.
The described robot has no external moving parts, which means that it might be silent, stealthy, reliable, cheap for manufacturing, and efficient.
It eliminates many pieces of the drivetrain with a directly driven propeller.

\begin{acknowledgments}
The authors would like to express gratitude to Profs. A.D.D. Craik, FRSE, T.J. Pedley, FRS, and H.K. Moffatt, FRS
for reading this manuscript and making useful comments.
Many thanks to Profs. M. Al-Ajmi, I.A. Eltayeb, D.W. Hughes, A.R. Kacimov, D. Kapanadze, and M.R.E. Proctor, FRS for helpful discussions.
This research is partially supported by the grant IG/SCI/ DOMS/18/16 from SQU, Oman.
\end{acknowledgments}

\end{document}